\documentclass[12pt, draftclsnofoot, onecolumn]{IEEEtran}
\usepackage{graphicx}
\ifCLASSINFOpdf
\else
\fi
\usepackage{url}
\usepackage{cite}
\usepackage{setspace}
\usepackage{amsmath}
\usepackage{amssymb}
\usepackage{bm}
\usepackage{nccmath}
\usepackage{ulem}
\usepackage{algorithm,algorithmic}
\usepackage{mathtools}
\usepackage{tikz}
\usepackage{tabularx,multirow,booktabs}

\usepackage{subcaption}

\usepackage{algorithm,algorithmic}

\begin{document}

\title{\LARGE Dimension Reduction-based Signal Compression for Uplink Distributed MIMO C-RAN with Limited Fronthaul Capacity}

\author{\normalsize \IEEEauthorblockN{Fred Wiffen\IEEEauthorrefmark{1}\IEEEauthorrefmark{3}, Mohammud Z. Bocus\IEEEauthorrefmark{2}, Woon Hau Chin\IEEEauthorrefmark{3}, Angela Doufexi\IEEEauthorrefmark{1} and Mark Beach\IEEEauthorrefmark{1}} \\
\IEEEauthorblockA{\IEEEauthorrefmark{1}Communication Systems \& Networks Group, University of Bristol,
Bristol, UK} \\
 \IEEEauthorblockA{\IEEEauthorrefmark{3}Bristol Research and Innovation Laboratory, Toshiba Europe Ltd., Bristol, UK} \\
\IEEEauthorblockA{\IEEEauthorrefmark{2}Ofcom, London, UK} \\
\IEEEauthorblockA{Email: \{fred.wiffen, angela.doufexi, mark.beach\}@bristol.ac.uk, zubeir.bocus@ofcom.org.uk, woonhau.chin@toshiba-bril.com}
}


\maketitle

\begin{abstract}
This paper proposes a dimension reduction-based signal compression scheme for uplink distributed MIMO cloud radio access networks (C-RAN) with an overall excess of receive antennas, in which users are jointly served by distributed multi-antenna receivers connected to a central processor via individual finite-capacity fronthaul links. We first show that, under quantization noise-limited operation, applying linear dimension reduction at each receiver before compressing locally with a uniform quantization noise level results in a sum capacity that scales approximately linearly with fronthaul capacity, and can come within a fixed gap of the cut-set bound. The dimension reduction filters that maximize joint mutual information are then shown to be truncated forms of the conditional Karhunen-Loeve transform, with a block coordinate ascent algorithm for finding a stationary point given. Analysis and numerical results indicate that the signal dimension can be reduced without significant loss of information, particularly at high signal-to-noise ratio, preserving the benefits of using excess antennas. The method is then adapted for the case of imperfect channel state information at the receivers. The scheme significantly outperforms conventional local signal compression at all fronthaul rates, and with complexity linear in network size represents a scalable solution for distributed MIMO C-RAN systems.
\end{abstract}


\IEEEpeerreviewmaketitle

\section{Introduction}
In a distributed multiple input multiple output (MIMO) uplink system, $K$ users are jointly served by $L$ geographically distributed receivers (or remote radio heads), each equipped with $M$ antennas. This distribution of receivers provides macro-diversity and improves uniformity of service, and is facilitated by the recent shift towards a cloud radio-access-network (C-RAN) architecture, in which processing for multiple receivers is performed at a single central processor (CP). A significant practical challenge with C-RAN MIMO, however, is the transfer of data from the receivers to the CP -- the large data rates associated with the transfer of raw IQ samples \cite{brubaker2016emerging}, combined with a growing interest in replacing fixed fibre with reduced capacity wireless point-to-point connections \cite{lombardi2018microwave} resulting in a need for efficient lossy compression of the received signals.

Here we propose a compression scheme for systems with an overall excess of receive antennas, $ML \gg K$, in which a dimension reduction filter is applied at each receiver to produce a reduced number, $N<M$, of signal components, which are then compressed locally at each receiver using simple transform coding. We show that by jointly designing the receive filters for all receivers, inter-receiver signal dependencies can be exploited to significantly reduce the number of signal components required to capture the received information, and that this compression method can efficiently utilise the available fronthaul.

\subsection{Related Work}
The challenge of signal compression for fronthaul constrained MIMO networks has received much research attention, see e.g. \cite{7444125}, \cite{7096298} and references therein. Significant attention has been given to compress-and-forward (CF) architectures, in which compression is applied to the received signal at each receiver before forwarding to the CP for decompression and symbol detection. Schemes in which each receiver independently compresses and forwards its own signal, e.g. \cite{6226311}, \cite{7009970} are attractive for their simplicity, and are currently implemented in practical systems. However, such schemes do not exploit the inherent dependencies between signals at different receivers and therefore do not operate efficiently at low fronthaul rates.

On the other hand, CF schemes based on distributed source (Wyner-Ziv) coding -- in which the signals from all receivers are jointly decompressed at the CP -- best exploit signal dependencies to achieve efficient compression \cite{el2011network}. The information theoretic limits of distributed source coding techniques for uplink MIMO systems have been studied in \cite{5238767}, and multiple compression schemes have been proposed. For example, in \cite{6824778} it is shown that under a total fronthaul capacity constraint a uniform quantization noise level across antennas achieves within a constant gap of the sum-capacity of the uplink C-RAN model, whilst \cite{6342931} shows that successive decompression can be achieved by transform coding with the conditional Karhunen-Loeve transform (KLT). However, the increased decoder complexity required for distributed source coding may be prohibitive for large networks with mobile channels.

Point-to-point compression schemes in which signals for each receiver are decompressed individually, but using compression codebooks that are jointly designed to exploit dependencies between receivers, present an attractive compromise. Under Gaussian signalling, the optimal point-to-point compression scheme can be implemented using transform coding. However, jointly finding the optimal transforms and rate allocations for each receiver involves numerically solving a non-convex optimization, using, for example, successive convex approximation \cite{7063645}, and has complexity that does not scale well with network size \cite{7465790}. In \cite{6824778} it is shown that at high signal-to-quantization-plus-noise ratio (SQNR) a uniform quantization noise (UQN) level approximately maximizes sum capacity. Other point-to-point compression methods in \cite{7134796} and \cite{8762078} reduce complexity by using locally calculated transforms followed by a centrally calculated rate allocation, aiming to maximize minimum user rate and sum capacity, respectively. An interesting observation from \cite{7134796} is that at lower fronthaul rates the optimal rate allocation is sparse -- only a subset of the $M$ signal components are quantized at each receiver. 

The sparse rate allocation in \cite{7134796} can be seen as an implicit dimension reduction, which is the basis of the method proposed in this paper. Dimension reduction is a common feature of signal compression schemes, and was previously applied to distributed wireless sensor networks in \cite{4276987}, with the aim of reducing the dimension of correlated sensor observations whilst minimizing the mean squared error of the signal estimate. Similarly the use of distributed compressive sensing for dimension reduction has also been investigated \cite{1600024}. In an uplink C-RAN context dimension reduction is explicitly performed at each receiver in \cite{8671721} using an analog beamforming stage before digital signal compression, whilst in \cite{7490884} compressive sensing is applied to the collated signals from all co-operating receivers to reduce the signal dimension prior to compression. The uplink dimension reduction concept has parallels to the widely studied downlink sparse beamforming approach \cite{6920005}, in which each transmitter only transmits to a subset of the users, thereby reducing the number of data streams transferred over fronthaul. 

\subsection{Contributions}
The main contributions of this paper are:
\begin{itemize}
	\item An analysis of UQN compression at high SNR is provided to show that reducing the signal dimension using a linear filter improves the rate at which system sum capacity increases with fronthaul capacity in the fronthaul-limited operating region.
    \item It is shown that the dimension reduction filters that maximize the joint mutual information between the transmit symbols and reduced dimension signals are a truncated form of the conditional Karhunen-Loeve transform. A block-coordinate ascent algorithm for finding a stationary point to the joint mutual information maximization is given, and it is argued that a large proportion of the full dimension information can be captured with significantly reduced signal dimension, particularly at high SNR.
    \item A simple modification to the proposed method is provided to account for imperfect channel state information under MMSE channel estimation.
    \item Numerical examples are given showing that dimension reduction significantly improves the rate-capacity performance of UQN compression, enabling operation close to the cut-set bound.
\end{itemize}

\subsection{Notation}
We use typeface $a$, $\mathbf{a}$ and $\mathbf{A}$ for scalars, column vectors and matrices respectively, and $\mathbf{A}^T$, $\mathbf{A}^*$, $\mathbf{A}^\dagger$ and $\det(\mathbf{A})$ respectively for the transpose, conjugate, conjugate transpose and determinant of $\mathbf{A}$. $\mathbf{I}_n$ represents the $n \times n$ identity matrix. Matrix inequalities, $\mathbf{A} \gg \mathbf{B}$, are taken elementwise. $\mathbb{E}\big[.\big]$ denotes the expectation operator.

\section{System Model}
\subsection{System Model}
We consider an uplink system in which $L$ distributed MIMO receivers, each equipped with $M$ antennas, jointly serve $K$ single antenna users. The receivers are connected via individual rate-constrained fronthaul links to a central processing unit (CPU), which uses signals from all of the receivers to jointly detect and decode the transmitted user symbols. We focus on the case where there is an overall excess of receive antennas, $ML \gg K$, such that the total number of signal observations in the network, $ML$, is much greater than the underlying signal dimensionality, $K$.

The received uplink signal at receiver $l$, $\mathbf{y}_l \in \mathbb{C}^{M}$, is given by
\begin{equation}
    \mathbf{y}_l = \mathbf{H}_l\mathbf{x} + \bm{\eta},
\end{equation}
where $\mathbf{H}_l \in \mathbb{C}^{M \times K}$ is the channel to receiver $l$, $\mathbf{x} \sim \mathcal{CN}\big(0,\rho\mathbf{I}_{K}\big)$ are independent Gaussian user uplink symbols with signal-to-noise ratio (SNR) $\rho$, and $\bm{\eta} \sim \mathcal{CN}\big(0,\mathbf{I}_{M}\big)$ additive white Gaussian noise with unit
variance.
The channel matrix $\mathbf{H}_l$ contains the power-control adjusted channel vectors
\begin{align}
    \mathbf{H}_l &= \bar{\mathbf{H}}_l\mathbf{P}^{1/2} \label{eqn:powercont} = \begin{bmatrix} \sqrt{p_1}\mathbf{h}_{l,1} & \ldots & \sqrt{p_K}\mathbf{h}_{l,K} \end{bmatrix},
\end{align}
where column $k$ of $\bar{\mathbf{H}}_l \in \mathbb{C}^{M \times K}$ is the propagation channel between user $k$ and receiver $l$, $\mathbf{h}_{l,k}$, and $\mathbf{P} \in \mathbb{C}^{K \times K}$ is a diagonal matrix containing the user power control coefficients, $p_k$. The channel matrix has eigenvalue decomposition,
\begin{equation}
    \mathbf{H}_l\mathbf{H}_l^\dagger = \mathbf{U}_l\mathbf{\Lambda}_l\mathbf{U}_l^\dagger
\end{equation}
where $\mathbf{U}_l \in \mathbb{C}^{M \times M}$ is a unitary matrix and $\mathbf{\Lambda}_l \in \mathbb{C}^{M \times M}$ a diagonal matrix containing the ordered channel eigenvalues, $\lambda_{l,i}$, of which $t$ are non-zero. We assume that the $\mathbf{H}_l$ are full rank, i.e. $t = \min(M,K)$,

Each receiver has digital processing capability and is connected to the CPU via an individual fronthaul connection with capacity $\mathcal{R}$ bits per channel use (bpcu). We assume that each receiver has access to its local CSI, and that all processing using global CSI is performed at the CP, which has access to CSI as appropriate.

Numerical results provided throughout this paper use the configuration set up described in Section \ref{sec:numres}.

\section{Reduced Dimension Signal Compression}
\label{sec:rationale}
We propose to compress the received signals in two stages. First, a linear dimension reduction filter, $\mathbf{W}_l \in \mathbb{C}^{M \times N}$, is applied at each receiver
\begin{equation}
\label{eqn:filter}
    \mathbf{z}_l = \mathbf{W}_l^\dagger\mathbf{y}_l,
\end{equation}
to produce signals with a reduced number, $N$, of dimensions, where $K/L \leq N \leq t$.

These reduced dimension signals are then quantized separately at each receiver using local optimal point-to-point block compression with a uniform quantization noise level\footnote{This should not be confused with fixed-rate scalar quantization with uniformly spaced quantization steps.} (UQN) to give quantized signal
\begin{equation}
    \tilde{\mathbf{z}}_l = \mathbf{z}_l + \boldsymbol{\delta}_l,
\end{equation}
where $\boldsymbol{\delta}_l \sim \mathcal{CN}\big(0,\Delta_l\mathbf{I}_N\big)$, with $\Delta_l$ the uniform quantization noise variance at receiver $l$.
\begin{figure}[t]
    \centering
    {\includegraphics[width=0.7\linewidth]{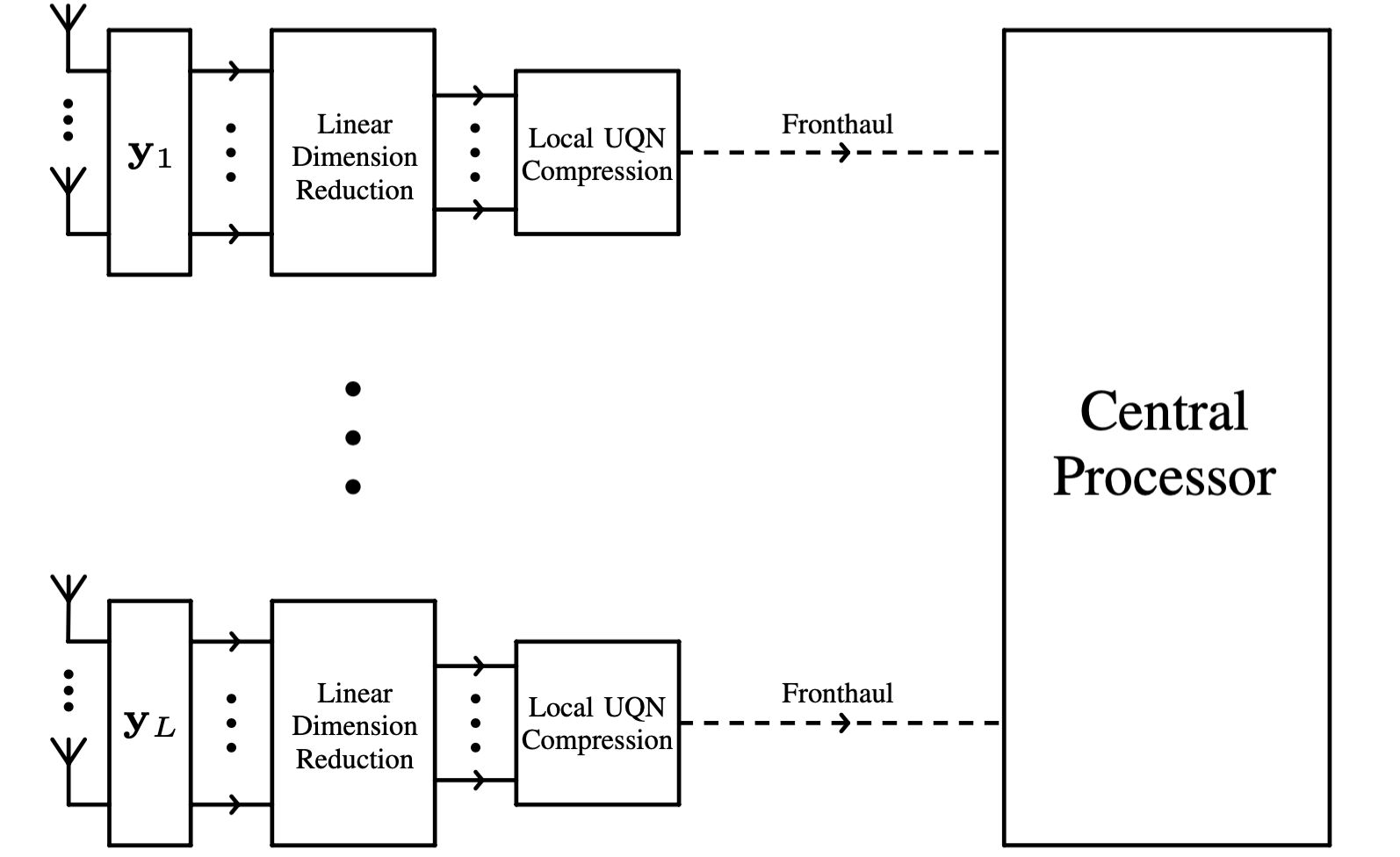}};
    \caption{Proposed C-RAN Signal Compression Architecture}
    \label{fig:dimreddiag}
\end{figure}
We now show that at high SNR, in the fronthaul limited region -- i.e. where system performance is limited by the total available fronthaul, $\mathcal{R}L$, rather than user transmit power -- the sum capacity of the MIMO C-RAN network scales like
\begin{equation}
\label{eqn:sumcapRK/N}
    \mathcal{C}_{\scriptscriptstyle \mathrm{SUM}} \approx \frac{\mathcal{R}K}{N} + \epsilon_{\scriptscriptstyle N},
\end{equation}
under either successive interference cancellation or linear symbol detection. Choosing the minimum signal dimension, $N = K/L$ (the total number of signal components is equal to the number of users), a sum user capacity can therefore be achieved that comes within a fixed gap of the total available fronthaul capacity
\begin{equation}
    \mathcal{C}_{\scriptscriptstyle \mathrm{SUM}} \approx \mathcal{R}L + \epsilon_{\scriptscriptstyle K/L},
\end{equation}
where $\epsilon_{\scriptscriptstyle K/L} \leq 0$.

\subsection{Reduced Dimension Channel}
Restricting our attention to filters with orthonormal columns (cf. Section \ref{sec:dimred}), the reduced dimension signals may be described by an equivalent channel
\begin{align}
    \mathbf{z}_l = \mathbf{G}_l\mathbf{x} + \tilde{\boldsymbol{\eta}},
\end{align}
where $\mathbf{G}_l = \mathbf{W}_l^\dagger\mathbf{H}_l$ and $\tilde{\boldsymbol{\eta}} \sim \mathcal{CN}\big(0,\mathbf{I}_N\big)$. The equivalent channel has eigenvalue decomposition
\begin{equation}
\label{eqn:locchaneig}
    \mathbf{G}_l\mathbf{G}_l^\dagger = \mathbf{V}_l\mathbf{\Gamma}_l\mathbf{V}_l^\dagger,
\end{equation}
where $\mathbf{V}_l$ contains the $N$ eigenvectors and $\mathbf{\Gamma}_l$ the $N$ ordered eigenvalues, $\gamma_{l,i}$, along its diagonal.

\subsection{UQN Compression}
The reduced dimension signal must be compressed
\begin{equation}
    \tilde{\mathbf{z}}_l = f_l\big(\mathbf{z}_l\big)
\end{equation}
at rate $\mathcal{I}\big(\tilde{\mathbf{z}}_l;\mathbf{z}_l\big) = \mathcal{R}$ for transfer over fronthaul.  The optimal compressed signal, under either a sum capacity or weighted user capacity measure, is given by the Gaussian vector test channel
\begin{equation}
    \tilde{\mathbf{z}}_l = \mathbf{z}_l + \boldsymbol{\delta}_l
\end{equation}
with independent quantization noise $\boldsymbol{\delta}_l \sim \mathcal{CN}\big(0,\mathbf{\Phi}_l\big)$. The optimal quantization noise covariance matrices, $\mathbf{\Phi}_l$, may be jointly found subject to rate constraints as a stationary point to a non-convex optimization problem. Unfortunately, this requires numerical methods based on majorization-minimization, which have high computational complexity that scales poorly with network size. 

The work in  \cite{6824778} shows that in the high signal-to-noise-plus-quantization ratio (SQNR) region, i.e. when $\rho\mathbf{G}_l\mathbf{G}_l^\dagger \gg \mathbf{I}_N + \boldsymbol{\Phi}_l$, a uniform quantization noise level (UQN), 
\begin{equation}
    \mathbf{\Phi}_l = \Delta_l\mathbf{I}_N,
\end{equation}
is asymptotically optimal (in terms of sum capacity). The uniform quantization noise level is independent for each receiver, and can be easily found by numerically solving
\begin{equation}
\label{eqn:fullsumcap}
    \mathcal{R} = \log_2 \det\Big(\mathbf{I}_N + \frac{\rho\mathbf{G}_l\mathbf{G}_l^\dagger + \mathbf{I}_N}{\Delta_l}\Big).
\end{equation}
Compression can then be performed using simple transform coding (cf. Section \ref{sec:transcod}). In the high SNR region, when $\rho\gamma_{l,i} \gg 1$ for all $i$, (\ref{eqn:fullsumcap}) may be approximated as
\begin{equation}
\label{eqn:approxrate}
    \mathcal{R} \approx \log_2 \det\Big(\frac{\rho\mathbf{G}_l\mathbf{G}_l^\dagger}{\Delta_l}\Big)  
\end{equation}
and the uniform quantization noise level is approximately
\begin{equation}
\label{eqn:quantnoiseapprox}
    \Delta_l \approx \rho \overline{\gamma}_l 2^{-\mathcal{R}/N}
\end{equation}
where $\overline{\gamma}_l = \big(\prod_i \gamma_{l,i} \big)^{1/N}$.

\subsection{System Capacity under UQN Compression}
\begin{figure}[h]
    \centering
    \includegraphics[width=0.7\linewidth]{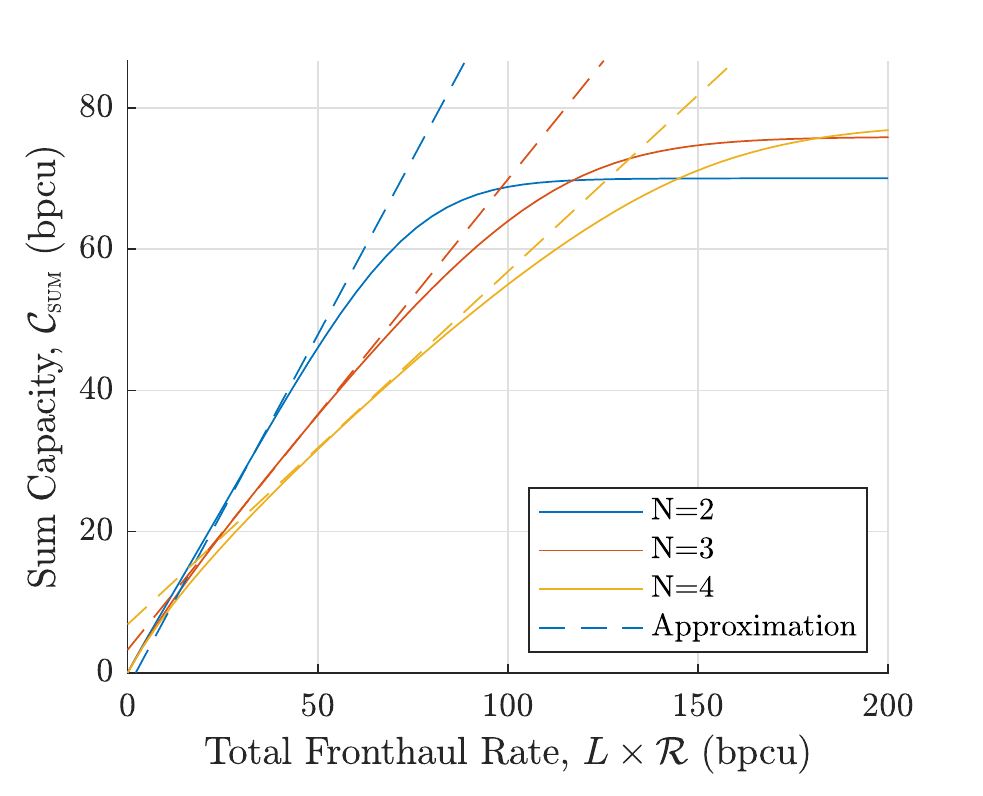}
    \caption{Rate-capacity curves for UQN compression under dimension reduction, $K=8$, $L=4$, $M = 8$, $\rho =15$ dB.}
    \label{fig:sumcapapprox}
\end{figure}

Under UQN compression it is straightforward to show that system sum capacity is given by
\begin{align}
    \mathcal{C}_{\scriptscriptstyle \mathrm{SUM}} &= \mathcal{I}\big(\tilde{\mathbf{z}}_1,\ldots,\tilde{\mathbf{z}}_L;\mathbf{x}\big) \\
    &= \log_2\det\Big(\mathbf{I}_K + \rho\sum_{l=1}^L\frac{\mathbf{G}_l\mathbf{G}_l^\dagger}{1 + \Delta_l}\Big).
\end{align}
In the fronthaul-limited operating region, i.e. when $\Delta_l \gg 1$, this can be approximated by
\begin{equation}
    \mathcal{C}_{\scriptscriptstyle \mathrm{SUM}} \approx \log_2\det\Big(\mathbf{I}_K + \rho\sum_{l=1}^L\frac{\mathbf{G}_l\mathbf{G}_l^\dagger}{\Delta_l}\Big).
\end{equation}
At all but very low fronthaul rates this can be further approximated by
\begin{equation}
    \mathcal{C}_{\scriptscriptstyle \mathrm{SUM}} \approx \log_2\det\Big( \rho\sum_{l=1}^L\frac{\mathbf{G}_l\mathbf{G}_l^\dagger}{\Delta_l}\Big).
\end{equation}
Applying the approximation in (\ref{eqn:quantnoiseapprox}),
\begin{align}
    \mathcal{C}_{\scriptscriptstyle \mathrm{SUM}} &\approx \log_2\det\Big( 2^{\mathcal{R}/N}\sum_{l=1}^L\frac{1}{\overline{\gamma}_l}\mathbf{G}_l\mathbf{G}_l^\dagger\Big) \\
    &= \frac{\mathcal{R}K}{N} + \log_2\det\Big(\sum_{l=1}^L\frac{1}{\overline{\gamma}_l}\mathbf{G}_l\mathbf{G}_l^\dagger\Big).
    \label{eqn:lincapapprox}
\end{align}
Hence we see that the sum capacity scales in inverse proportion to the reduced signal dimension, with an offset that is independent of fronthaul rate and SNR\footnote{These approximations are only tight when $\rho\gamma_{l,i} \gg 1$, however numerical results indicate that the general trend of improved scaling with reduced signal dimension holds at lower SNR.}.  This is illustrated in Figure \ref{fig:sumcapapprox}.

This scaling occurs when capacity is limited by quantization noise, which scales with $2^{-\mathcal{R}/N}$. Considering (\ref{eqn:lincapapprox}) it therefore makes sense to use a small signal dimension when fronthaul capacity is limited. The constraint $N \geq K/L$ ensures that the number of signal observations at the CP is at least the number of transmitted symbols. However, since reducing the signal dimension discards information, when quantization noise is similar or smaller than receiver noise the use of a larger signal dimension can provide higher capacity, as discussed in Section \ref{sec:achievablerates}.

With no dimension reduction applied, the sum capacity can be approximated (at high SNR)
\begin{align}
    \mathcal{C}_{\scriptscriptstyle \mathrm{SUM}} &\approx \frac{\mathcal{R}K}{t} + \log_2\det\Big(\sum_{l=1}^L\frac{1}{\overline{\lambda}_l}\mathbf{H}_l\mathbf{H}_l^\dagger\Big),
\end{align}
and therefore for systems with an excess of receive antennas, $ML \gg K$, conventional UQN compression is inefficient in the fronthaul-limited region (since $\mathcal{R}K/t \ll \mathcal{R}L$). This contrasts with the high SQNR region, where conventional UQN compression performs well.

In principle this improved scaling can be achieved at high signal to noise ratios using any dimension reduction method -- for example simply reducing the number of antennas or selecting a subset at each receiver -- but not all dimension reduction methods will achieve the same performance. In the next section we show that the use of jointly designed receive filters in a system with an excess of antennas enables a good description of the received information with a relatively small number of signal components.

\section{Dimension Reduction Filter Design}
\label{sec:dimred}
At each receiver a dimension reduction filter is applied to the received signal as in (\ref{eqn:filter}), to give
\begin{equation}
    \mathbf{z}_l = \mathbf{W}_l^\dagger\mathbf{H}_l\mathbf{x} + \tilde{\boldsymbol{\eta}}_l,
\end{equation}
with filtered noise $\tilde{\boldsymbol{\eta}}_l \sim \mathcal{CN}\big(0,\mathbf{W}_l^\dagger\mathbf{W}_l\big)$. Without loss of generality, we may restrict our attention to receive filters with orthonormal columns\footnote{By the (thin) QR decomposition, any $\bar{\mathbf{W}} \in \mathbb{C}^{M \times N}$ with independent columns may be written $\bar{\mathbf{W}} = \mathbf{W}\mathbf{R}$, where $\mathbf{\mathbf{W}} \in \mathbb{C}^{M \times N}$ has orthonormal columns and $\mathbf{R} \in \mathbb{C}^{N \times N}$ is upper triangular. Since $\mathbf{R}$ is invertible it does not affect the information captured by $\bar{\mathbf{W}}^\dagger\mathbf{y}$.}
\begin{equation}
\label{eqn:orthonorm}
    \mathbf{W}_l^\dagger\mathbf{W}_l = \mathbf{I}_N.
\end{equation}
Reducing the signal dimension to $N < t$ necessarily incurs a loss of information. We choose to find the set of filters, $\mathbf{W}_1,\ldots,\mathbf{W}_L$, that maximize the joint mutual information between the uncompressed reduced dimension signals and the user symbols,
\begin{equation}
    \underset{\mathbf{W}_1,\ldots,\mathbf{W}_L}{\mathrm{maximize}} \quad \mathcal{I}(\mathbf{z}_1,\ldots,\mathbf{z}_L;\mathbf{x}).
\end{equation}
This objective function captures the dependencies between signals at different receivers, and intuitively acts as a good heuristic for maximising overall sum capacity\footnote{Different objective functions that explicitly account for individual user service requirements may be an area for future work.}. Furthermore, as we will see, this formulation leads to a simple iterative algorithm for finding the filters, which we show are truncated forms of the conditional Karhunen-Loeve transform -- previously used in other distributed signal compression applications \cite{1194019}. 

For illustration, we first consider the case of dimension reduction for a single receiver, before deriving the iterative algorithm for jointly finding the receive filters for multiple receivers. We show that, at high SNR, the resulting filters are independent of user power control, and, extending to multi-antenna users, independent of user transmit beamforming. We then argue, using analytical and numerical results, that this dimension reduction can be performed without significant loss of information. 

\subsection{Single Receiver Dimension Reduction}
\label{sec:single}
The mutual information between the reduced dimension signal at receiver $l$ and the transmit symbols is
\begin{equation}
\label{eqn:singlemax}
    \mathcal{I}(\mathbf{z}_l ; \mathbf{x}) = \log_2 \det \big(\mathbf{I}_{N} + \rho\mathbf{W}_l^\dagger\mathbf{H}_l\mathbf{H}_l^\dagger\mathbf{W}_l\big).
\end{equation}
It is shown in Appendix 1 that, under the orthonormality constraint, a global maximum to this is achieved by setting the columns of $\mathbf{W}_l$ to be equal to the $N$ principal eigenvectors of $\mathbf{H}_l\mathbf{H}_l^\dagger$ (those corresponding to the $N$ largest eigenvalues). This is equivalent to taking the $N$ principal components of $\mathbf{y}_l$, or the first $N$ outputs of the classical Karhunen-Loeve transform (KLT). We therefore refer to this dimension reduction filter as the truncated KLT (T-KLT). 

The amount of information captured \textit{locally} under T-KLT dimension reduction is a function of the local channel eigenvalues
\begin{equation}
\label{eqn:localMI}
    \mathcal{I}(\mathbf{z}_l ; \mathbf{x}) = \sum_{i=1}^{N} \log_2\big(1 + \rho \lambda_{l,i}\big),
\end{equation}
as is the amount of information lost
\begin{equation}
    \mathcal{I}\big(\mathbf{y}_l;\mathbf{x}\vert \mathbf{z}_l\big) = \sum_{i=N+1}^{t} \log_2\big(1 + \rho \lambda_{l,i}\big).
\end{equation}
Whilst a large eigenvalue spread is generally linked to poor MIMO performance, a large eigenvalue spread in the \textit{local} distributed MIMO channels is attractive from a dimension reduction perspective, as it enables a high proportion of information to be captured by a small number of signal components. In a distributed MIMO system, where local channel strengths vary and power control is applied at a global level, larger local eigenvalue spreads are an inherent feature.


Since the signals at all receivers are dependent, the information loss at a \textit{global} level is reduced 
\begin{equation*}
    \mathcal{I}\big(\mathbf{y}_l;\mathbf{x}\vert \mathbf{z}_1,\ldots,\mathbf{z}_L\big) \leq \sum_{i=N+1}^{t} \log_2\big(1 + \rho \lambda_{l,i}\big),
\end{equation*}
which indicates the potential to apply dimension reduction more aggressively in a multi-receiver network. 

However, since the T-KLT depends on only local CSI and does not account for inter-receiver signal dependencies, it generally does not maximize the joint mutual information in a multi-receiver setting. We now extend the filter design method to maximize joint mutual information.

\subsection{Multiple Receiver Dimension Reduction}
\label{sec:t-cklt}
We wish to find the $\mathbf{W}_l$ that jointly capture the maximum information about $\mathbf{x}$, given by
\begin{equation}
\label{eqn:cap1}
    \mathcal{I}(\mathbf{z}_1,\ldots,\mathbf{z}_L;\mathbf{x}) = \log_2\det \big( \mathbf{I}_K + \rho\sum_{l=1}^L\mathbf{H}_l^\dagger\mathbf{W}_l\mathbf{W}_l^\dagger\mathbf{H}_l\big)
\end{equation}
i.e. we wish to solve
\begin{equation}
\label{eqn:123rd}
\begin{aligned}
& \underset{\mathbf{W}_1,\ldots,\mathbf{W}_L}{\mathrm{maximize}} \quad
\det \big( \mathbf{I}_K + \rho\sum_{l=1}^L\mathbf{H}_l^\dagger\mathbf{W}_l\mathbf{W}_l^\dagger\mathbf{H}_l\big)  \\
& \mathrm{subject \ to} \quad \quad \mathbf{W}_l^\dagger\mathbf{W}_l = \mathbf{I}_{N}. 
\end{aligned}
\end{equation}
Whilst this problem is non-convex in the $\mathbf{W}_l$, a stationary point may be found using a simple block coordinate ascent algorithm. First, note that (\ref{eqn:cap1}) may be expanded in terms of the conditional mutual information,
\begin{equation}
 \label{eqn:condexp}
     \mathcal{I}(\mathbf{z}_1,\ldots,\mathbf{z}_L;\mathbf{x}) = \mathcal{I}(\mathbf{z}_l;\mathbf{x}\vert \mathbf{z}_{l}^{\mathsf{c}}) + \mathcal{I}(\mathbf{z}_{l}^{\mathsf{c}};\mathbf{x}),
\end{equation}
where $\mathbf{z}_{l}^{\mathsf{c}} = [\mathbf{z}_{1};\ldots;\mathbf{z}_{l-1};\mathbf{z}_{l+1};\ldots;\mathbf{z}_{L}]$. In Appendix 2 it is shown that
\begin{equation}
    \mathcal{I}(\mathbf{z}_l;\mathbf{x}\vert \mathbf{z}_{l}^{\mathsf{c}}) = \log_2 \det \big( \mathbf{I}_{N} + \rho\mathbf{W}_l^\dagger\mathbf{H}_l\mathbf{A}_{l}\mathbf{H}_l^\dagger\mathbf{W}_l \big),
\end{equation}
where
\begin{equation}
    \mathbf{A}_{l} = \big(\mathbf{I}_{K} + \rho\sum_{i\neq l}\mathbf{H}_i^\dagger\mathbf{W}_i\mathbf{W}_i^\dagger\mathbf{H}_i\big)^{-1}.
\end{equation}
Since $\mathcal{I}(\mathbf{z}_{l}^{\mathsf{c}};\mathbf{x})$ is independent of $\mathbf{W}_l$, for fixed $\mathbf{W}_1,\ldots,\mathbf{W}_{l-1},\mathbf{W}_{l+1},\ldots,\mathbf{W}_{L}$ the optimal filter at receiver $l$ is found by solving
\begin{equation}
\label{eqn:global_max}
\begin{aligned}
& \underset{\mathbf{W}_l}{\text{maximize}} \quad
\det \big( \mathbf{I}_{N} + \rho\mathbf{W}_l^\dagger\mathbf{H}_l\mathbf{A}_{l}\mathbf{H}_l^\dagger\mathbf{W}_l \big)  \\
& \mathrm{subject \ to} \quad \quad \mathbf{W}_l^\dagger\mathbf{W}_l = \mathbf{I}_{N} 
\end{aligned}
\end{equation}
Similarly to (\ref{eqn:singlemax}), this is achieved by setting the columns of $\mathbf{W}_l$ to be the $N$ principal eigenvectors of $\mathbf{H}_l\mathbf{A}_{l}\mathbf{H}_l^\dagger$. Noting that 
\begin{equation}
    \mathbb{E}\big[\mathbf{y}_l\mathbf{y}_l^\dagger \big| \mathbf{z}_{l}^{\mathsf{c}}\big] = \mathbf{I}_{M} + \rho\mathbf{H}_l\mathbf{A}_{l}\mathbf{H}_l^\dagger
\end{equation}
we observe that the optimal filter is a truncated form of the `conditional' KLT \cite{4016296} (T-CKLT herein), where conditioning is with respect to the other reduced dimension signals. This makes intuitive sense -- the filter columns at each receiver correspond to the $N$ `directions' in which there is the largest uncertainty given the reduced dimension observations provided by the other receivers.\footnote{Note that the mutual information maximizing dimension reduction filters differ to the MMSE dimension reduction filters derived in \cite{4276987}.}

A stationary point to (\ref{eqn:123rd}) may accordingly be found using a block coordinate ascent (BCA) procedure \cite{5285192} \cite{1664999}, iteratively updating the $\mathbf{W}_l$ in turn, as shown in Algorithm \ref{alg:optfilter}. Here the $\mathbf{W}_l$ are initialised using the T-KLT filters. 
\begin{algorithm}
\caption{T-CKLT block-coordinate ascent (BCA) dimension reduction filter design}
\label{alg:optfilter}
\begin{algorithmic}
\STATE \textbf{inputs:} $\mathbf{H}_l \quad \forall l$ 
\vspace{5px}
\STATE $\mathbf{W}_l \gets$ $N$ principal eigenvectors of $\mathbf{H}_l\mathbf{H}_l^\dagger \quad \forall l$ \vspace{3px}\\ 
\FOR {$j = 1 : j_{\mathrm{max}} $}
    \FOR {$l = 1 : L $}
        \STATE {\vspace{3px}
        $\mathbf{A}_l \gets \big(\mathbf{I}_{K} + \rho\sum_{i\neq l}\mathbf{H}_i^\dagger\mathbf{W}_i\mathbf{W}_i^\dagger\mathbf{H}_i\big)^{-1}$ \\ \vspace{5px}
        $\mathbf{W}_l \gets$ $N$ principal eigenvectors of $\mathbf{H}_l\mathbf{A}_l\mathbf{H}_l^\dagger$ \vspace{3px}}
    \ENDFOR
\ENDFOR
\STATE \textbf{outputs:} $\mathbf{W}_{l} \quad \forall l$
\end{algorithmic}
\end{algorithm}

\begin{figure}[h]
    \centering
    \includegraphics[width=0.7\linewidth]{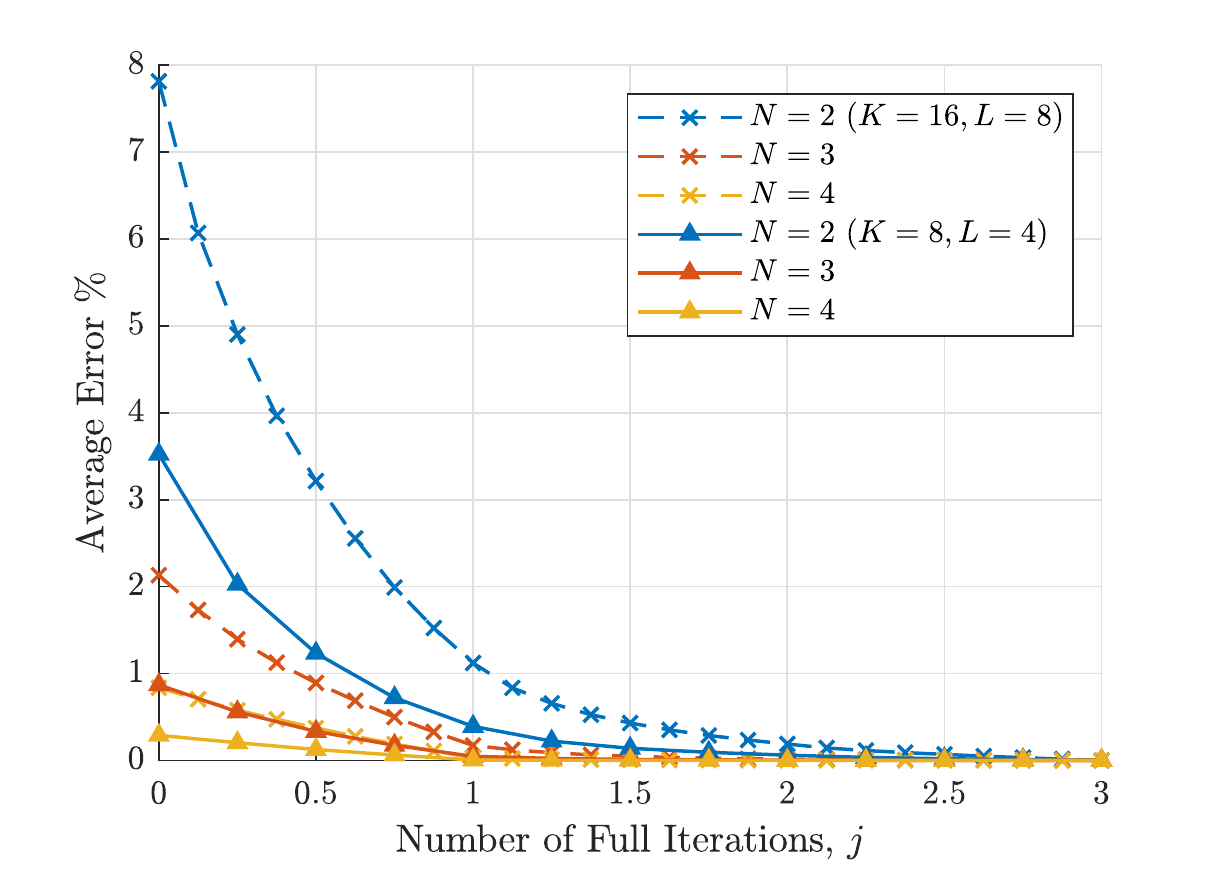}
    \caption{Convergence of T-CKLT BCA algorithm, $M = 8$, $\rho = 15$ dB.}
    \label{fig:convergence}
\end{figure}

At each iteration $\mathcal{I}(\mathbf{z}_1,\ldots,\mathbf{z}_L;\mathbf{x})$ monotonically increases, and hence Algorithm \ref{alg:optfilter} converges to a stationary point of (\ref{eqn:123rd}). Simulations indicate that a small number of iterations ($j_{\mathrm{max}} \leq 3$) are typically required to converge to within a practical tolerance of the maximum, as shown in Figure \ref{fig:convergence}.

\subsection{Dimension Reduction at High SNR}
At high $\rho$, we have
\begin{equation}
    \lim_{\rho \to \infty} \ \rho\mathbf{H}_l\mathbf{A}_l\mathbf{H}_l^\dagger = \mathbf{H}_l\Big(\sum_{i \neq l}\mathbf{H}_i^\dagger\mathbf{W}_i\mathbf{W}_i^\dagger\mathbf{H}_i\Big)^{-1}\mathbf{H}_l^\dagger
\end{equation}
and the maximum joint mutual information dimension reduction filters become independent of SNR. Furthermore, substituting in (\ref{eqn:powercont}),
\begin{equation}
    \mathbf{H}_l\Big(\sum_{i \neq l}\mathbf{H}_i^\dagger\mathbf{W}_i\mathbf{W}_i^\dagger\mathbf{H}_i\Big)^{-1}\mathbf{H}_l^\dagger = \bar{\mathbf{H}}_l\Big(\sum_{i \neq l}\bar{\mathbf{H}}_i^\dagger\mathbf{W}_i\mathbf{W}_i^\dagger\bar{\mathbf{H}}_i\Big)^{-1}\bar{\mathbf{H}}_l^\dagger,
\end{equation}
the filters are independent of the individual user power control coefficients -- decoupling dimension reduction filter design from user power control. 

\subsection{Dimension Reduction with Multi-Antenna Users}
Extending the dimension reduction method to the case of multi-antenna terminals each with $c$ antennas, we may write
\begin{equation}
    \mathbf{y}_l = \bar{\mathbf{H}}_l\mathbf{\Theta}\mathbf{x} + \boldsymbol{\eta},
\end{equation}
where $\bar{\mathbf{H}}_l = \begin{bmatrix} \bar{\mathbf{H}}_{l,1} & \ldots & \bar{\mathbf{H}}_{l,K} \end{bmatrix}$ contains the user propagation channel matrices, $\bar{\mathbf{H}}_{l,k} \in \mathbb{C}^{M \times c}$, and $\mathbf{\Theta} = \mathrm{diag}(\mathbf{\Theta}_k)$ is a block diagonal matrix containing user transmit beamforming matrices, $\mathbf{\Theta}_k \in \mathbb{C}^{c \times c}$. Assuming all user beamforming matrices are full-rank, at high SNR the receive filters are given, as above, by the principal eigenvectors of
\begin{equation}
\bar{\mathbf{H}}_l\mathbf{\Theta}\big(\sum_{i\neq l}\mathbf{\Theta}^\dagger\bar{\mathbf{H}}_i^\dagger\mathbf{W}_i\mathbf{W}_i^\dagger\bar{\mathbf{H}}_i\mathbf{\Theta}\big)^{-1}\mathbf{\Theta}^\dagger\bar{\mathbf{H}}_l^\dagger =
\bar{\mathbf{H}}_l\big(\sum_{i\neq l}\bar{\mathbf{H}}_i^\dagger\mathbf{W}_i\mathbf{W}_i^\dagger\bar{\mathbf{H}}_i\big)^{-1}\bar{\mathbf{H}}_l^\dagger.
\end{equation}
This shows that, under joint mutual information maximization, dimension reduction filters can be chosen independently of user beamforming matrices.

\subsection{Analytical Performance Insights}
\label{sec:dranalysis}
We now use some simple asymptotic results to give insight into the behaviour of the filters.
The joint mutual information can be expressed
\begin{align}
    \mathcal{I}\big(\mathbf{z}_1,\ldots,\mathbf{z}_L;\mathbf{x}\big) &= \log_2\det\Big(\mathbf{I}_K + \rho\sum_{l=1}^L\mathbf{G}_l\mathbf{G}_l^\dagger\Big) \\
    &= \sum_{k=1}^K \log_2\big(1 + \rho\upsilon_k\big)
\label{eq:eqeig}
\end{align}
where $\upsilon_k$ are the eigenvalues of the \textit{global} equivalent channel, $\sum_l\mathbf{G}_l^\dagger\mathbf{G}_l$. For practical operation it is required that this global equivalent channel be full rank, i.e. all $\upsilon_k > 0$, implying that $LN\geq K$ -- the total number of signal components transferred to the CPU is greater than the number of users. Since the mutual information scales linearly with the number of non-zero eigenvalues but only logarithmically with their magnitudes, we can expect the set of receive filters that maximize (\ref{eq:eqeig}) to generally produce a full rank channel whenever $LN\geq K$. This is consistent with all simulations carried out.

The total information lost due to the dimension reduction operation is
\begin{equation}
    \mathcal{L} = \mathcal{I}\big(\mathbf{y}_1,\ldots,\mathbf{y}_L;\mathbf{x}\vert\mathbf{z}_1,\ldots,\mathbf{z}_L\big),
\end{equation}
to which the T-CKLT BCA algorithm achieves a minima. Defining a matrix $\overline{\mathbf{W}}_l \in \mathbb{C}^{M \times (M-N)}$ such that $\overline{\mathbf{W}}_l^{\dagger}\mathbf{y}_l$ is the signal component discarded during dimension reduction, i.e $\mathbf{W}_l^\dagger\overline{\mathbf{W}}_l = \mathbf{0}$, and $\overline{\mathbf{G}}_l = \overline{\mathbf{W}}_l^\dagger\mathbf{H}_l$, then it can be shown that
\begin{equation}
\mathcal{L} =\log_2\det\Big(\mathbf{I}_K + \rho \sum_{l=1}^L\overline{\mathbf{G}}_l^\dagger\overline{\mathbf{G}}_l \Big(\mathbf{I}_K + \rho\sum_{i=1}^L \mathbf{G}_i^\dagger\mathbf{G}_i\Big)^{-1}\Big).
\end{equation}
Assuming full rank global channel this may be upper bounded
\begin{equation}
    \mathcal{L} < \log_2\det\Big(\mathbf{I}_K +  \sum_{l=1}^L\overline{\mathbf{G}}_l^\dagger\overline{\mathbf{G}}_l \Big(\sum_{i=1}^L \mathbf{G}_i^\dagger\mathbf{G}_i\Big)^{-1}\Big)
\end{equation}
which is tight when $\rho\sum_{l=1}^L \mathbf{G}_l^\dagger\mathbf{G}_l \gg \mathbf{I}_K$. This bound is independent of $\rho$, implying that at high SNR dimension reduction causes a constant information loss that depends only on the channel and signal dimension $N$. In contrast, from (\ref{eqn:cap1}) the joint mutual information increases with $\rho$, and as a result the proportion of information lost due to dimension reduction vanishes as $\rho \to \infty$. This holds for any dimension reduction filter that produces a full rank channel. 

\begin{figure}[h]
    \centering
    \includegraphics[width=0.7\linewidth]{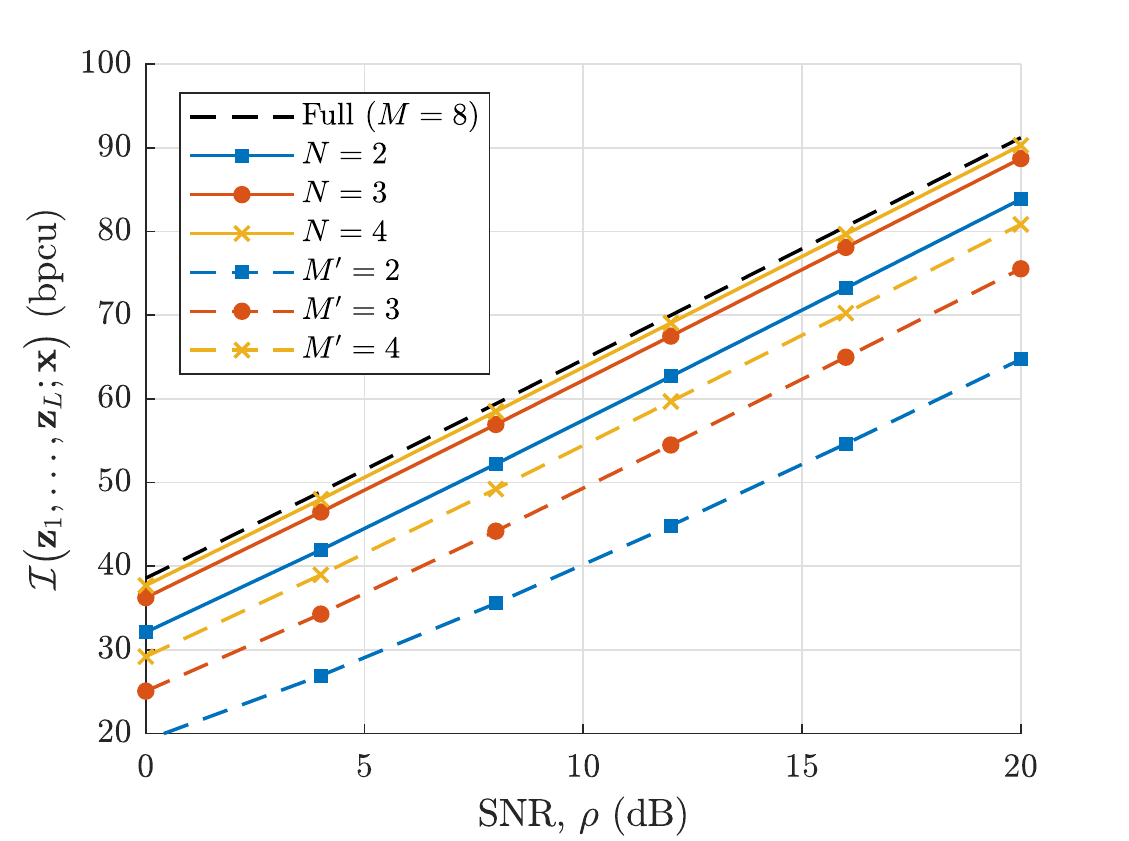}
    \caption{Scaling of $\mathcal{I}\big(\mathbf{z}_1,\ldots,\mathbf{z}_L;\mathbf{x}\big)$ with $\rho$, with $K=8$, $L=4$, $M = 8$.}
    \label{fig:capsnr}
\end{figure}

Figure \ref{fig:capsnr} shows the benefit of the proposed method compared to simply reducing the number of antennas ($M'$), with the T-CKLT filters capturing most of the information in the full signal.

In maximizing the conditional mutual information at each stage, the T-CKLT BCA method inherently produces signals with reduced redundancy between receivers. Intuitively, from a compression perspective this is attractive -- if the signals at the distributed receivers have low redundancy or dependencies, then simple local signal compression applied at each receiver can be expected to perform well, with little benefit to using more complex compression schemes.

\section{Signal Compression}
\label{sec:sigcomp}
After dimension reduction, UQN compression is performed separately at each receiver to give
\begin{equation}
\label{eqn:quantz}
    \tilde{\mathbf{z}}_l = \mathbf{z}_l + \boldsymbol{\delta}_l
\end{equation}
where $\boldsymbol{\delta}_l \sim \mathcal{CN}(0,\Delta_l)$, with $\Delta_l$ chosen such that
\begin{equation}
\label{eqn:locquant}
    \mathcal{R} =\log_2 \det\Big(\mathbf{I}_N + \frac{\rho\mathbf{G}_l\mathbf{G}_l^\dagger + \mathbf{I}_N}{\Delta_l}\Big).
\end{equation}
This can be solved for $\Delta_l$ using a simple bisection algorithm and implemented using transform coding.

\subsection{Transform Coding}
\label{sec:transcod}
Using the eigendecomposition (\ref{eqn:locchaneig}), (\ref{eqn:locquant}) may be written as
\begin{equation}
    \mathcal{R} = \sum_{i=1}^N \log_2\Big(1 + \frac{\rho\gamma_{l,i} + 1}{\Delta_l}\Big)
\end{equation}
which we note is the sum of $N$ scalar Gaussian compression channels with rates
\begin{equation}
\label{eqn:scalrat}
    r_{l,i} = \log_2\Big(1 + \frac{\rho\gamma_{l,i} + 1}{\Delta_l}\Big).
\end{equation}
The uniform quantization noise level can therefore be achieved using a transform coding approach, consisting of:
\begin{enumerate}
    \item a linear locally decorrelating transform\footnote{A uniform quantization noise level can be achieved by direct scalar compression of the components of $\mathbf{z}_l$ without decorrelating transform, but results in a higher quantization noise level than given by (\ref{eqn:locquant}).}, $\mathbf{V}_l^\dagger\mathbf{z}_l$, to produce $N$ (locally) independent scalars, with variances $\sigma_{l,i}^2 = \rho\gamma_{l,i} + 1$.
    \item individual Gaussian scalar compression of the $N$ scalars using quantization rates $r_{l,i}$, to give quantization noise level $\Delta_l$.
    \item a reverse transform, $\mathbf{V}_l$, at the CP to produce $\tilde{\mathbf{z}}_l$.
\end{enumerate}
When all $\rho\gamma_{l,i}  \gg 1$ and $2^{\mathcal{R}/N} \gg 1$, the quantization noise level is approximately given by
\begin{equation}
\label{eqn:uniformquantlevel}
    \Delta_l \approx \rho \overline{\gamma}_l 2^{-\mathcal{R}/N}
\end{equation}
with $\overline{\gamma}_l = \big(\prod_{i = 1}^{N}\gamma_{l,i}\big)^{1/N}$. Compared to the full dimension case, the $\rho\gamma_{l,i}  \gg 1$ condition will tend to be satisfied at lower SNR, $\rho$, since the dimension reduction filter will tend to produce reduced dimension channels with $N$ strong eigenvalues.

To avoid numerically solving for $\Delta_l$ the approximation
\begin{equation}
    r_{l,i} \approx \log_2\Big(\frac{\rho\gamma_{l,i}}{\Delta_l}\Big)
\end{equation}
can be used, and it follows that the rate allocation
\begin{equation}
\label{eqn:ratealloc}
    r_{l,i} = \frac{\mathcal{R}}{N} + \log_2(\gamma_{l,i}) - \frac{1}{N}\sum_{j=1}^N\log_2(\gamma_{l,j})
\end{equation}
approximately achieves the uniform quantization noise level in (\ref{eqn:uniformquantlevel}). Transform coding with this rate allocation in fact maximizes local mutual information, $\mathcal{I}\big(\tilde{\mathbf{z}}_l;\mathbf{x}\big)$, as discussed in \cite{6875354}.

\subsection{Practical Quantizers}
Optimal Gaussian scalar compression as considered in the analysis here is an information theoretic concept that assumes the use of infinite length coding blocks, and thus cannot be implemented in a real system. Practical schemes using spherical or trellis codes have been shown to come close to this performance using short block lengths \cite{hamkins2002gaussian}, whilst simple fixed-rate Lloyd-Max scalar quantization achieves the same noise level using an additional $1.4$ bits per signal component \cite{952802}. Since the number of signal components is small, fixed-rate quantization represents an attractive low complexity compromise for practical systems.

\section{Achievable Uplink Capacity}
\label{sec:achievablerates}
As above, the sum capacity, under optimal successive interference cancellation detection is, of the proposed scheme is
\begin{equation}
\label{eqn:sumcap1}
    \mathcal{C}_{\scriptscriptstyle \mathrm{SUM}} = \log_2\det\Big(\mathbf{I}_K + \rho\sum_{l=1}^L\frac{\mathbf{G}_l\mathbf{G}_l^\dagger}{1 + \Delta_l}\Big),
\end{equation}
and in the fronthaul limited region ($\Delta_l \gg 1$) can be approximated.
\begin{equation}
    \mathcal{C}_{\scriptscriptstyle \mathrm{SUM}} \approx \frac{\mathcal{R}K}{N} + \log_2\det\Big(\sum_{l=1}^L\frac{1}{\bar{\gamma}_l}\mathbf{G}_l\mathbf{G}_l^\dagger\Big).
\end{equation}
On the other hand, when quantization noise power becomes small compared to noise, $\Delta_l \ll 1$, the sum capacity is approximately
\begin{equation}
    \mathcal{C}_{\scriptscriptstyle \mathrm{SUM}} \approx \log_2\det\Big(\mathbf{I}_K + \rho\sum_{l=1}^L\mathbf{G}_l\mathbf{G}_l^\dagger\Big),
\end{equation}
and the information loss due to dimension reduction becomes the limiting factor.

At low fronthaul rates it is therefore desirable to use small $N$, but as the available rate increases and the system moves into the noise limited region $N$ must be increased to increase capacity. The maximum sum capacity at a given rate $\mathcal{R}$ can be found by calculating $\mathbf{W}_l$ and $\Delta_l$ for different values of $N$ and directly comparing (\ref{eqn:sumcap1}). The overall rate-capacity curve follows the envelope of the curves in Figure \ref{fig:sumcapapprox}, and is non-smooth.
 
\subsection*{Achievable User Capacities under LMMSE Detection}
The processing required to achieve the sum capacity in (\ref{eqn:sumcap1}) becomes prohibitive for larger systems and therefore lower complexity linear methods, such as linear minimum mean square error (LMMSE) symbol detection are often employed instead.

It can be shown that the LMMSE symbol estimates,
\begin{equation}
    \hat{\mathbf{x}} = \sum_{l=1}^L \mathbf{B}_l\tilde{\mathbf{z}}_l,
\end{equation}
are achieved using detection matrices
\begin{equation}
    \mathbf{B}_l =  \rho \Big(\mathbf{I}_K + \rho\sum_{i=1}^L\frac{\mathbf{G}_i^\dagger\mathbf{G}_i}{1 + \Delta_i}\Big)^{-1}\frac{\mathbf{G}_l^\dagger}{1 + \Delta_l}.
\end{equation}
The achievable rate of user $k$ is
\begin{equation}
    \mathcal{C}_k = \log_2\big(1 + \mathrm{SQINR}_k\big),
\end{equation}
where the signal-to-quantization-plus-interference-plus-noise ratio (SQINR) is
\begin{equation}
    \mathrm{SQINR}_k = \dfrac{1}{\Big[\Big(\mathbf{I}_K + \rho\displaystyle \sum_{l=1}^L\dfrac{\mathbf{G}_l^\dagger\mathbf{G}_l}{1 + \Delta_l}\Big)^{-1}\Big]_{k,k}} - 1.
\end{equation}
In the fronthaul limited region, following the same set of approximations above, i.e.
\begin{equation}
     \mathbf{I}_K + \rho\sum_{l=1}^L\frac{\mathbf{G}_l^\dagger\mathbf{G}_l}{1 + \Delta_l} \approx \rho\sum_{l=1}^L\frac{\mathbf{G}_l^\dagger\mathbf{G}_l}{\Delta_l}, \\
\end{equation}
the user capacities can be approximated 
\begin{equation}
    \mathcal{C}_k \approx \frac{\mathcal{R}}{N} - \log_2\Big(\Big[\Big(\displaystyle \sum_{l=1}^L\dfrac{\mathbf{G}_l^\dagger\mathbf{G}_l}{\overline{\gamma}_l}\Big)^{-1}\Big]_{k,k}\Big),
\end{equation}
and thus the sum user capacity under linear detection, $\sum_{k=1}^K \mathcal{C}_k$, also scales with $\mathcal{R}K/N$.

\section{Dimension Reduction Compression with Imperfect CSI}
\label{sec:imperfectcsi}
In practical systems, generally only noisy estimates of the user channels are available. Assuming MMSE channel estimation is used, the dimension reduction method is easily adapted for the case of imperfect CSI at the receivers. 

The channel estimate errors under MMSE estimation are
\begin{equation}
\label{eqn:lmmsechannelvec}
    \mathbf{e}_{l,k} = \mathbf{h}_{l,k} - \hat{\mathbf{h}}_{l,k},
\end{equation}
where $\hat{\mathbf{h}}_{l,k}$ is the estimated channel. For a given channel realisation and estimate, $\mathbf{e}_{l,k}$ is fixed (and unknown), but for randomly varying channels can be treated as a random variable with covariance $\mathbb{E}\big[\mathbf{e}_{l,k} \mathbf{e}_{l,k}^\dagger\big] = \mathbf{C}_{l,k}$, which, by the orthogonality principle, is uncorrelated with the channel estimate $\mathbb{E}\big[\hat{\mathbf{h}}_{l,k}\mathbf{e}_{l,k}^\dagger \big] = \mathbf{0}$. 

From (\ref{eqn:lmmsechannelvec}), the received signal at receiver $l$ may be expressed in terms of components through the known and unknown channel,
\begin{equation}
\label{eqn:ulchannerr}
\begin{aligned}
    \mathbf{y}_l =\hat{\mathbf{H}}_l\mathbf{x} + \mathbf{E}_l\mathbf{x} + \boldsymbol{\eta},
\end{aligned}
\end{equation}
with MIMO channel matrix estimate $\hat{\mathbf{H}}_l \in \mathbb{C}^{M \times K}$ and error $\mathbf{E}_l\in \mathbb{C}^{M \times K}$.

\subsection{Receive Filter Design}
Following the method in \cite{1193803}, the signal component through the unknown channel can be treated as noise
\begin{equation}
    \mathbf{y}_l = \hat{\mathbf{H}}_l\mathbf{x} + \boldsymbol{\omega}_l
\end{equation}
with $\boldsymbol{\omega}_l = \mathbf{E}_l\mathbf{x} + \boldsymbol{\eta}$. For a given channel estimate, $\mathbf{E}_l$ is unknown, and the statistics of $\boldsymbol{\omega}_l$ are therefore unknown. However, over random channel estimates $\boldsymbol{\omega}_l$ is uncorrelated with the known signal component, with covariance 
\begin{equation}
\label{eqn:intplusnoisecov}
    \begin{aligned}
        \mathbb{E}\big[\boldsymbol{\omega}_l\boldsymbol{\omega}_l^\dagger\big] &= \mathbf{I}_M + \rho\mathbf{C}_{l} \\
        &= \boldsymbol{\Omega}_l,
    \end{aligned}
\end{equation}
where the expectation is with respect to all quantities, and
\begin{equation}
    \mathbf{C}_{l} = \sum_{k=1}^K\mathbf{C}_{l,k}.
\end{equation}
A transform may be applied to whiten this channel estimation error noise,
\begin{equation}
    \begin{aligned}
    \check{\mathbf{y}}_l &= \boldsymbol{\Omega}_l^{-1/2}\mathbf{y}_l = \check{\mathbf{H}}_l\mathbf{x} + \check{\boldsymbol{\omega}}_l
    \end{aligned}
\end{equation}
where $\check{\mathbf{H}}_l = \boldsymbol{\Omega}_l^{-1/2}\hat{\mathbf{H}}_l$ and $\mathbb{E}\big[\check{\boldsymbol{\omega}}_l\check{\boldsymbol{\omega}}_l^\dagger\big] = \mathbf{I}_M$. Dimension reduction is then applied to this whitened signal
\begin{equation}
    \mathbf{z}_l = \mathbf{W}_l^\dagger\check{\mathbf{y}}_l.
\end{equation}
Following the reasoning of \cite{1193803}, the expected mutual information (with respect to the channel estimate error) is lower bounded
\begin{equation}
\label{eqn:capineq}
    \mathbb{E}\big[\mathcal{I}\big(\mathbf{z}_1,\ldots,\mathbf{z}_L;\mathbf{x} \vert\hat{\mathbf{H}}_l\big)\big] \geq \mathbb{E}\Big[\log_2 \det \Big(\mathbf{I}_K + \rho\sum_{l=1}^L\check{\mathbf{H}}_l^\dagger\mathbf{W}_l\mathbf{W}_l^\dagger\check{\mathbf{H}}_l \Big)\Big],
\end{equation}
with equality if the channel error is Gaussian. Following the same reasoning as Section \ref{sec:t-cklt}, applying the T-CKLT BCA algorithm to the transformed channels, $\check{\mathbf{H}}_l$, maximizes this lower bound. For spatially correlated user fading channels, this $\boldsymbol{\Omega}_l^{-1/2}$ transformation of the channel can be seen as a weighting which causes the T-CKLT BCA algorithm to favour signal subspaces which (on average) contain less channel estimation error. 

Since both desired signal and equivalent noise scale with $\rho$, the joint mutual information is bounded independently of $\rho$ 
\begin{equation}
    \log_2 \det \Big(\mathbf{I}_K + \rho\sum_{l=1}^L\check{\mathbf{H}}_l^\dagger\mathbf{W}_l\mathbf{W}_l^\dagger\check{\mathbf{H}}_l \Big) \\ < \log_2 \det \Big(\mathbf{I}_K + \sum_{l=1}^L\hat{\mathbf{H}}_l^\dagger\mathbf{C}_{l}^{-1/2}\mathbf{W}_l\mathbf{W}_l^\dagger\mathbf{C}_{l}^{-1/2}\hat{\mathbf{H}}_l \Big).
\end{equation}
In contrast with the perfect CSI case the reduced dimension signal captures a fixed proportion of the full dimension joint mutual information as $\rho \to \infty$ (since both are limited by channel errors, rather than transmit power). As the quality of the channel estimates increases ($\mathbf{C}_{l}$ decreases) the reduced dimension signal captures a greater proportion of available mutual information.


As before, a reduced dimension channel can be defined
\begin{equation}
    \hat{\mathbf{G}}_l = \mathbf{W}_l^\dagger\boldsymbol{\Omega}_l^{-1/2}\hat{\mathbf{H}}_l,
\end{equation}
with eigenvectors $\hat{\mathbf{V}}_l$ and eigenvalues $\hat{\gamma}_{l,i}$ as in (\ref{eqn:locchaneig}).

\subsection{Signal Compression}
Since the statistics of $\mathbf{z}_l$ are unknown (for a given channel estimate), a uniform quantization noise level cannot be perfectly achieved. However, the transform coding method above may be applied heuristically, using the estimated eigenvectors and eigenvalues, and expected to perform similarly when CSI errors are small.

The eigenvectors $\hat{\mathbf{V}}_l$ produce a decorrelated signal \textit{on average}, but for a given channel estimate do not produce perfectly independent signal components. Rate allocation is then performed using
\begin{equation}
    r_{l,i} = \frac{\mathcal{R}}{N} + \log_2(\hat{\gamma}_{l,i}) - \frac{1}{N}\sum_{j=1}^N\log_2(\hat{\gamma}_{l,j}).
\end{equation}
However, the variances of the $N$ scalars being quantized, 
\begin{equation}
\begin{aligned}
    \sigma_{l,i}^2 &= \big[\hat{\mathbf{V}}_l^\dagger\mathbb{E}\big[\mathbf{z}_l\mathbf{z}_l\big]\hat{\mathbf{V}_l}\big]_{i,i} \\
    &\neq \rho\hat{\gamma}_{l,i} + 1,
\end{aligned}
\end{equation}
are now not perfectly known, due to the uncertainty in the channel. For analytical tractability, we assume here that these variances \textit{are} perfectly known, which is a reasonable simplification, since a small mismatch in input variance to the quantizers can be tolerated with negligible performance loss \cite{1092819}. The resulting quantization noise covariance matrix is
\begin{equation}
    \mathbf{\Phi}_l = \mathbf{V}_l\Big(\mathrm{diag}\Big(\frac{\sigma_{l,i}^2}{2^{r_{l,i}} - 1}\Big)\Big)\mathbf{V}_l^\dagger,
\end{equation}
and compressed signals available to the CPU for symbol detection are
\begin{equation}
    \tilde{\mathbf{z}}_l = \hat{\mathbf{G}}_l\mathbf{x} + \boldsymbol{\delta}_l.
\end{equation}

\subsection{Achievable Sum Capacity}
Following the method of \cite{1193803}, the expected sum capacity can be lower bounded as
\begin{equation}
\label{eqn:capimperfectCSI}
    \mathbb{E}\big[\mathcal{C}_{\scriptscriptstyle \mathrm{SUM}}^{\scriptscriptstyle (\mathrm{CSI})}\big] \geq \mathbb{E}\Big[\log_2 \det \Big(\mathbf{I}_K + \rho\sum_{l=1}^L\hat{\mathbf{G}}_l^\dagger\big(\mathbf{I} + \mathbf{\Phi}_l\big)^{-1}\hat{\mathbf{G}}_l\Big)\Big].
\end{equation}

\section{Computational Complexity \& Signalling Overheads}
We now give a brief overview of the computation and signalling overheads associated with the proposed scheme.

\subsection{Computational Complexity}
Each iteration of the T-CKLT BCA algorithm requires an eigendecomposition with complexity $\mathcal{O}(M^3)$ and matrix inversion with complexity $\mathcal{O}(K^3)$. The transform coding signal compression requires a further eigendecomposition with complexity $\mathcal{O}(N^3)$ per receiver. The overall complexity at the CP\footnote{Simulations indicate that $j_{max}$ does not scale significantly with $L$} is therefore $\mathcal{O}\big((j_{max}(K^3 + M^3) + M^3 + N^3)L\big)$. With complexity linear in $L$, the proposed compression scheme, unlike optimal point-to-point compression -- which has complexity at least $\mathcal{O}\big(L^{3.5}\big)$ \cite{7465790}, scales well to larger network sizes.

The dimension reduction filter and decorrelating transform may be combined into a single matrix, $\mathbf{W}_l\mathbf{V}_l$, and communicated back to the receivers, and hence no significant computation is required at the receivers.

\subsection{Signalling}
The CP requires full CSI of $MK$ values per receiver, whilst the receivers require feedback of the decorrelated dimension reduction filters $\mathbf{W}_l\mathbf{V}_l$, $MN$ values. Assuming channel estimation is performed locally at the receivers, a total of $M(K+N)n_b$ bits of overhead must be communicated over fronthaul once per coherence block (assuming $n_b$ bits per complex value). We assume here that the coherence block is large, so that these overheads represent a small proportion of the total data payload. However, for highly mobile channels this may not be the case -- a more thorough consideration of this is left as future work.

\section{Numerical Results}
\label{sec:numres}
\begin{figure}[h]
    \centering
    \includegraphics[width=0.6\linewidth]{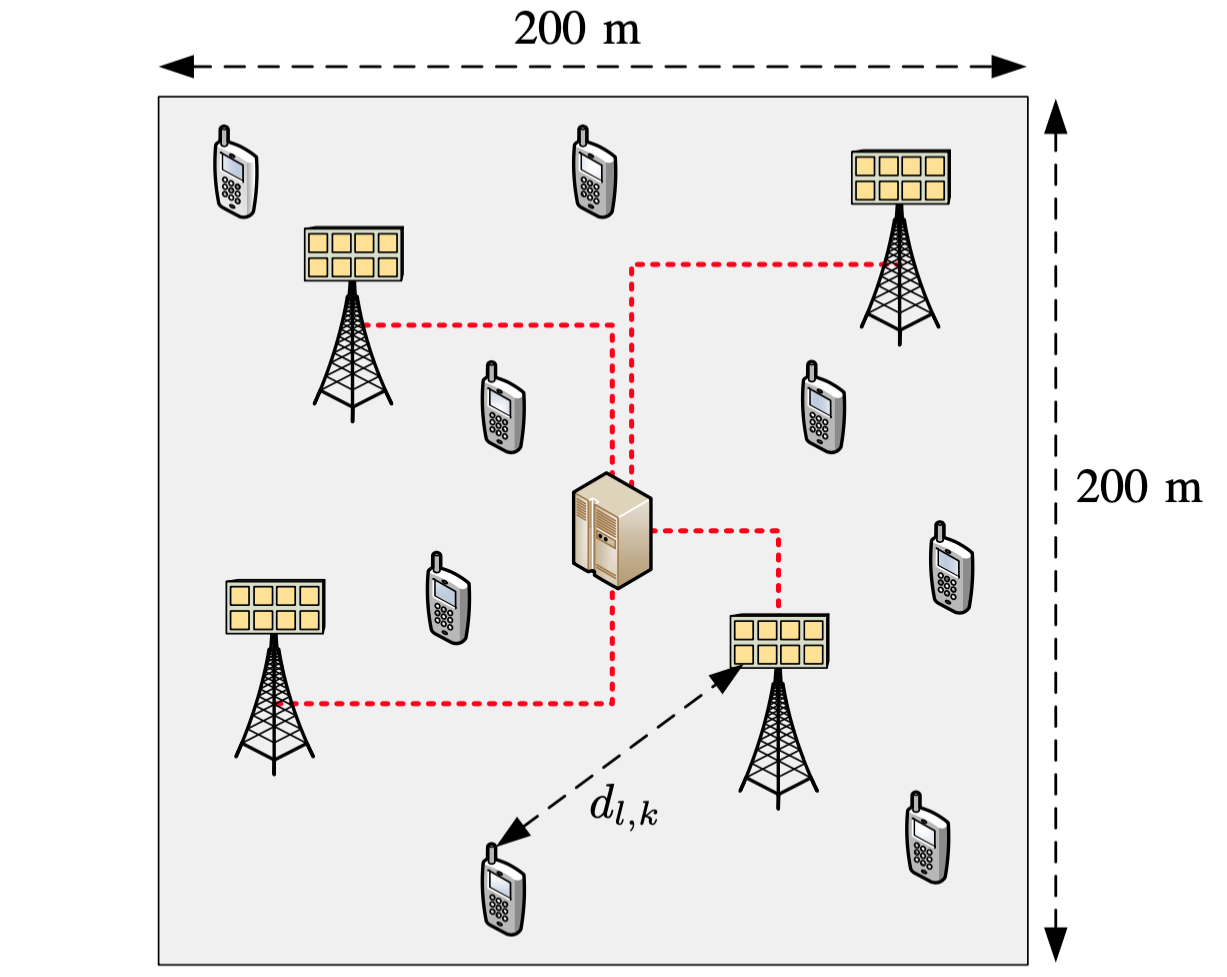}
    \caption{Illustration of system configuration with $K=8$, $L=4$, $M = 8$.}
    \label{fig:simconfig}
\end{figure}

We now present numerical results demonstrating the performance benefits of the dimension reduction compression approach, using a dense single-cell environment configuration in which the receivers and users are distributed randomly within a 200m $\times$ 200m area, as shown in Figure \ref{fig:simconfig}. 
Users are at a height of 1 m and receivers 6 m, with the channels following independent Rayleigh fading distributions,
\begin{equation}
    \mathbf{h}_{l,k} \sim \mathcal{CN}(0,\beta_{l,k}\mathbf{I}_M) \nonumber
\end{equation}
where $\beta_{l,k}$ follows a log-distance shadow fading path loss model with pathloss exponent 2.9 and 5.7 dB log-normal shadow fading \cite{7434656}. Power control, $p_k$, is applied such that the total average received power for each user over the whole network is the same
\begin{equation}
    \frac{1}{ML}\mathbb{E}\Big[\sum_{l=1}^L p_k\big \Vert \mathbf{h}_{l,k} \big \Vert^2  \Big] =  p_k\sum_{l=1}^L\frac{\beta_{l,k}}{L} = 1\nonumber.
\end{equation}

\subsection{Achievable Rates}
\begin{figure}
   \centering
        \includegraphics[width=0.7\columnwidth]{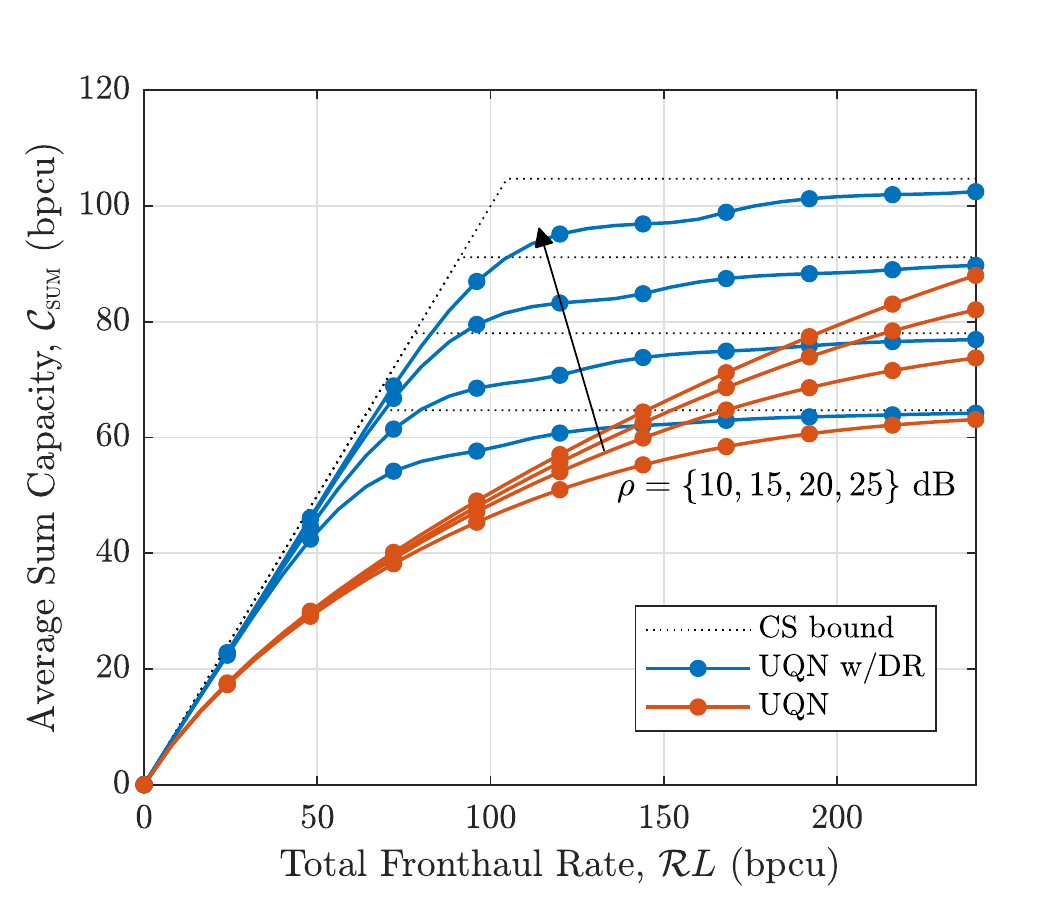}
    \caption{Rate-capacity performance of proposed scheme with $K=8,L=4,M=8,\rho=15$ dB, optimal $N$.}
    \label{fig:DR_cap_snr}
\end{figure}

Figure \ref{fig:DR_cap_snr} shows the average rate-capacity performance of UQN compression with and without dimension reduction, at different signal to noise ratios, for $K=8,L=4,M=8$. The sum capacity under dimension reduction is found by maximising over $N$ at each $\mathcal{R}$ (with the same value of $N$ for all receivers). 

Dimension reduction enables a substantial improvement in sum capacity at all values of $\mathcal{R}$, operating close to the cut-set bound in the fronthaul-limited region. The benefit is most pronounced at high SNR -- with $\rho = 25$ dB and a total fronthaul rate of 100 bpcu, a sum capacity of almost 90 bpcu can be achieved, compared to only 50 bpcu with standard UQN compression. The good performance at high $\rho$ follows from the fact that at high SNR only a small number of signal components at each receiver are required to capture most of the total received information. Whilst standard UQN is approximately optimal at higher fronthaul rates, the use of a dimension reduction stage offers a capacity improvement in all regions.

\begin{figure}
   \centering
        \includegraphics[width=0.65\columnwidth]{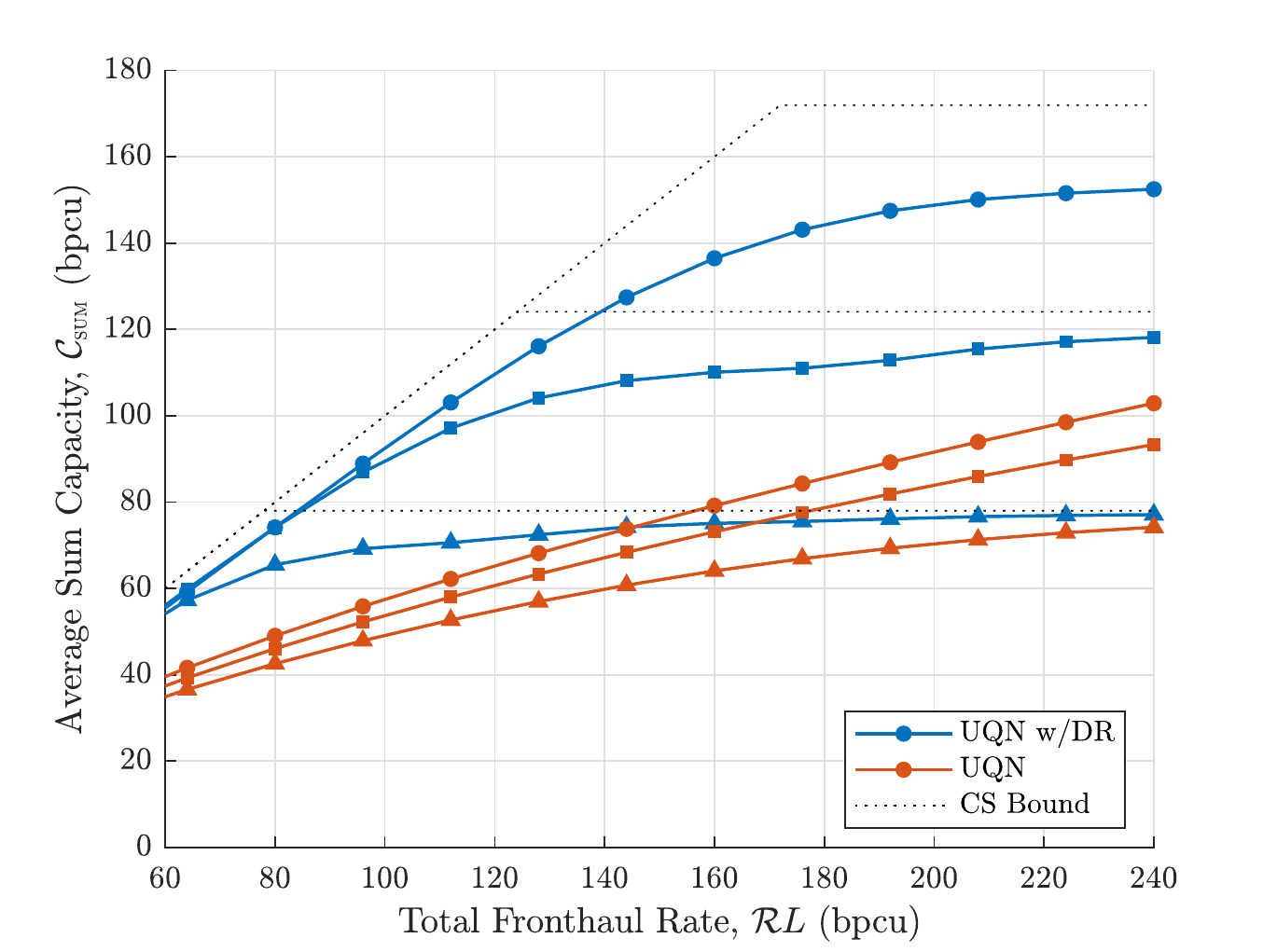}
    \caption{Rate-capacity performance of proposed scheme with varying user and receiver density, $M=8,\rho=15$ dB. Triangle markers:
    $K=8,L=4$, square markers: $K=12,L=6$, circle markers: $K=16,L=8$}
    \label{fig:compvarLvarK}
\end{figure}

These performance benefits continue as the density of users and receivers is increased, as shown in Figure \ref{fig:compvarLvarK}.

\subsection{Achievable User Rates under Linear Detection}
\begin{figure}
   \centering
        \includegraphics[width=0.65\columnwidth]{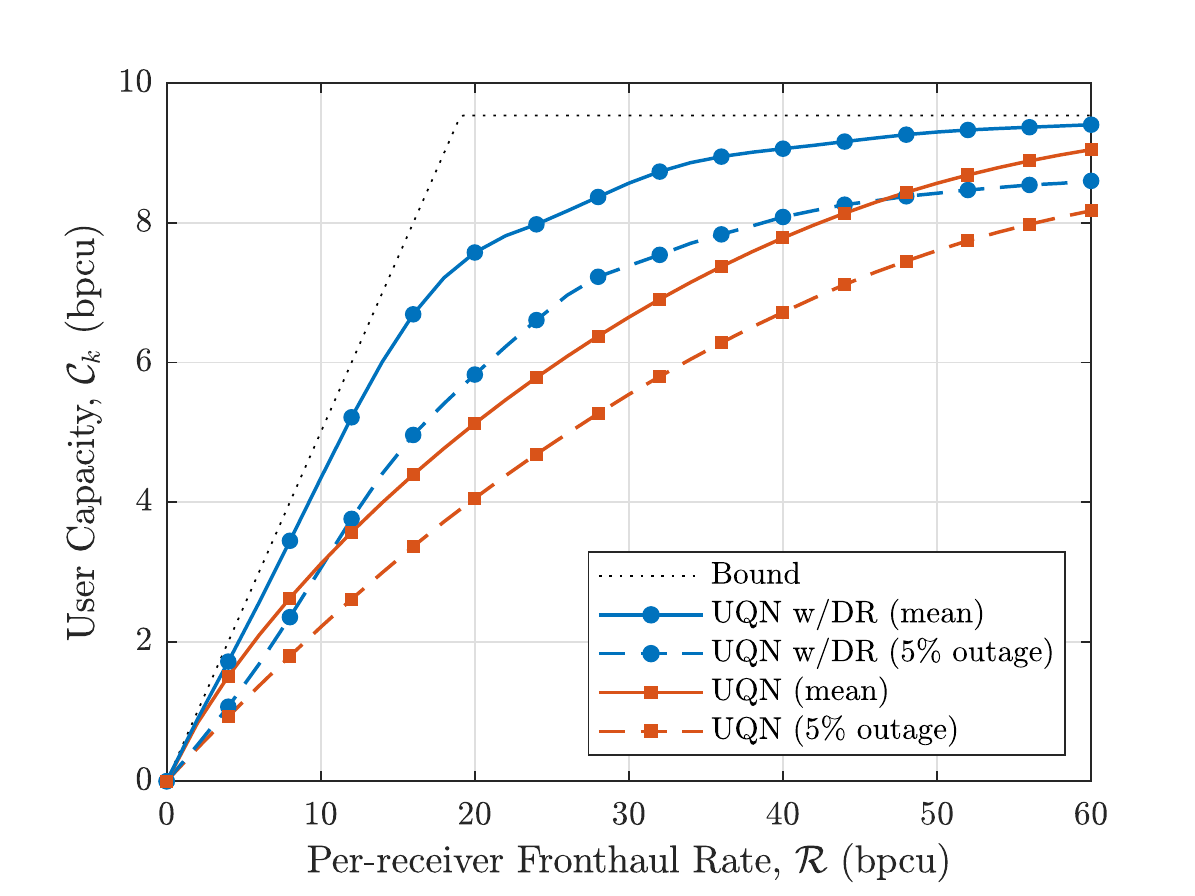}
    \caption{Rate-capacity performance of proposed scheme under linear detection, $K=8,L=4,M=8,\rho=15$ dB.}
    \label{fig:linear}
\end{figure}

Numerical results indicate that the proposed scheme can achieve similar performance benefits under linear symbol detection. Figure \ref{fig:linear} shows the average and 5\% outage user capacities, which are upper bounded by $\mathcal{R}L/K$ and the unquantized full dimension average and 5\% outage capacities, respectively. With a receiver fronthaul rate of $20$ bpcu, user mean and outage capacities are both improved by around 2 bpcu.

\subsection{Comparison of Dimension Reduction Methods}
\begin{figure}[h]
   \centering
        \includegraphics[width=0.7\columnwidth]{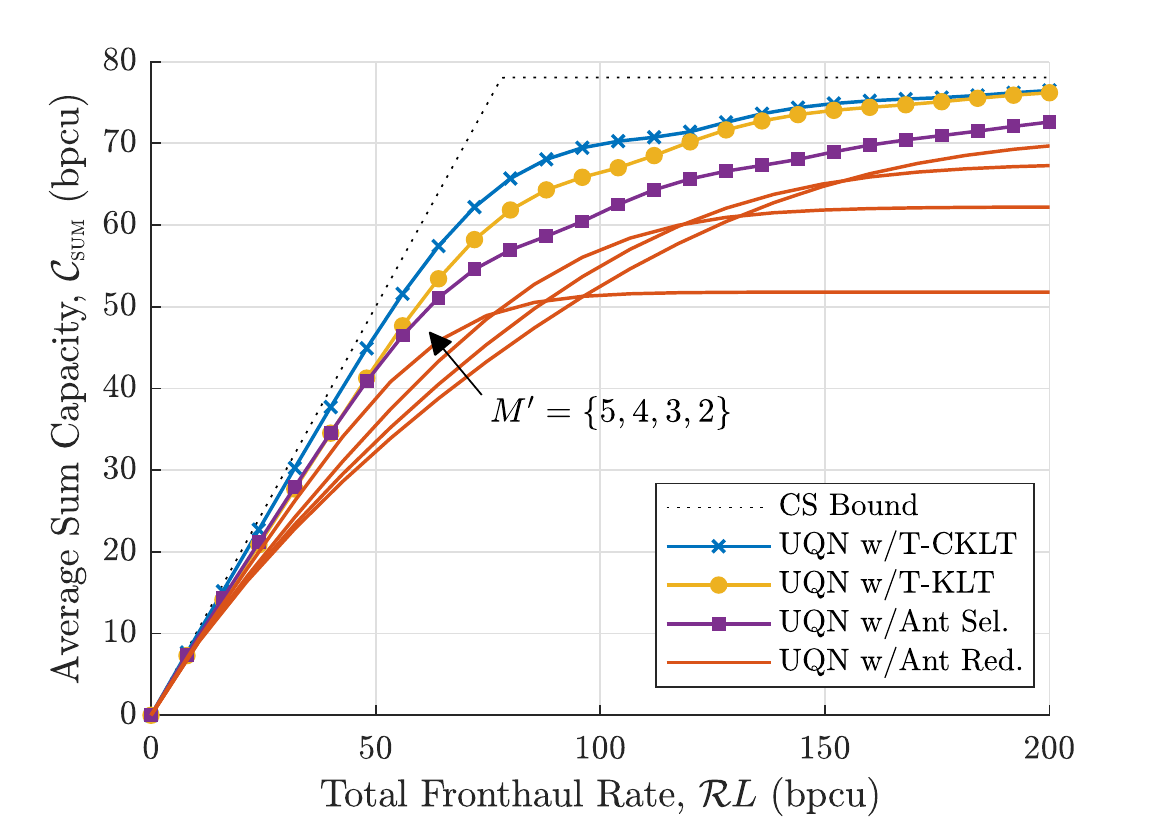}
    \caption{Rate-capacity performance of different dimension reduction schemes, $K=8,L=4,M=8,\rho=15$ dB.}
    \label{fig:alt_dr_comp}
\end{figure}

Following the discussion in Section \ref{sec:rationale}, we recall that any dimension reduction method can offer a potential capacity scaling improvement under fronthaul compression. Figure \ref{fig:alt_dr_comp} compares the performance of the T-CKLT dimension reduction scheme with T-KLT dimension reduction, an antenna selection based dimension reduction scheme in which $N$ out of $M$ antennas are selected at each receiver based on global CSI using a method adapted from \cite{1687757}, and simple reduction of the number of antennas.
The dimension reduction methods that exploit the diversity provided by the excess of antennas (T-CKLT, T-KLT and antenna selection) all provide a performance improvement compared to simply reducing the number of antennas, since they are able to exploit channel knowledge to provide an improved reduced dimension signal. This indicates that with limited fronthaul capacity there is still a benefit to deploying additional antennas. 

Whilst the T-KLT does not account for inter-receiver dependencies, its energy compaction properties enable it to achieve similar performance to the T-CKLT method, particularly at higher fronthaul rates. The T-KLT depends only on local CSI and may be calculated locally at the receivers, in which case only the reduced dimension equivalent channels, $\mathbf{G}_l$, need to be transferred to the CP. It therefore represents an attractive option for reducing signalling overheads and computational burden at the CP.

\subsection{Achievable Rates under Imperfect CSI}
\begin{figure}[h]
   \centering
        \includegraphics[width=0.7\columnwidth]{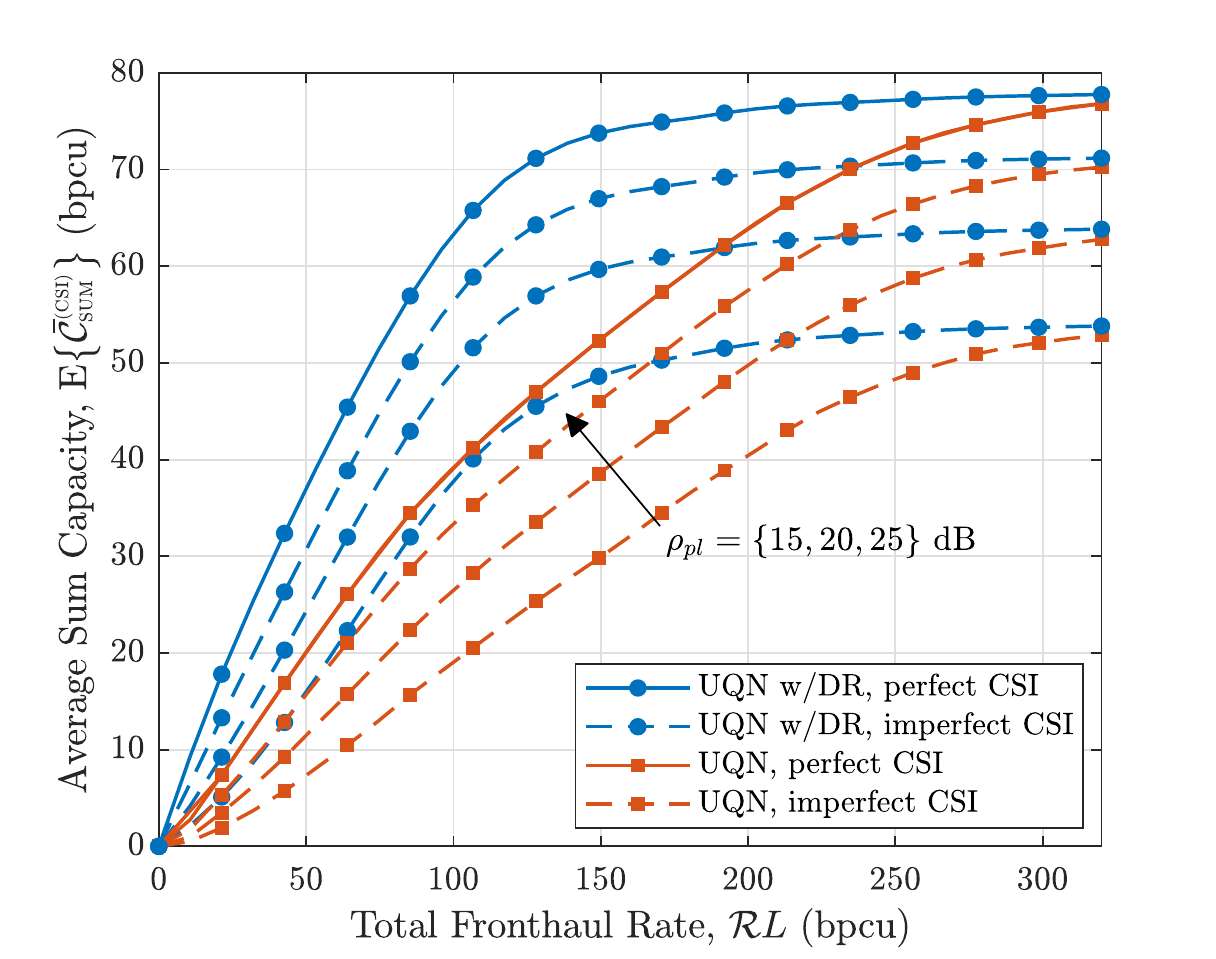}
    \caption{Rate-capacity performance of proposed scheme with varying quality of CSI, $K=8,L=4,M=8,\rho=15$ dB.}
    \label{fig:imperfect_CSI}
\end{figure}

Figure \ref{fig:imperfect_CSI} compares the rate-capacity performance of the proposed scheme in Section \ref{sec:imperfectcsi}, with and without the dimension reduction stage. The use of orthogonal pilots with SNR $\rho_{\mathrm{pl}}$ is assumed for channel estimation. For all pilot SNR levels a significant improvement in capacity is achieved through applying dimension reduction, with a fronthaul rate penalty incurred relative to the perfect CSI case.

\section{Conclusions}
In this paper we have shown that the application of dimension reduction to received signals enables simple local compression to achieve a system sum capacity that scales well with the total available fronthaul capacity. Algorithms for finding appropriate dimension reduction filters and implementing signal compression have been outlined for the cases of both perfect and imperfect CSI. Numerical results show that the scheme significantly outperforms conventional local compression, and can operate close to the cut-set capacity bound.

\section*{Acknowledgement}
This work was supported by the Engineering and Physical Sciences Research Council grant number EP/I028153/1 and Toshiba Europe Limited.

\bibliography{bibliography.bib}
\bibliographystyle{ieeetr}

\appendix

\subsection*{Appendix 1}
Consider the matrix product
\begin{equation}
    \mathbf{A}^\dagger\mathbf{B}\mathbf{A}
\end{equation}
where $\mathbf{B} \in \mathbb{C}^{n \times n}$ is a Hermitian symmetric matrix with ordered eigenvalues $\beta_i$, and $\mathbf{A} \in \mathbb{C}^{m \times n}$, $m \leq n$, a rectangular matrix with orthonormal columns, $\mathbf{A}^\dagger\mathbf{A} = \mathbf{I}_{m}$. By the Poincar\'e separation theorem \cite{bellman1997introduction}, the eigenvalues of $\mathbf{A}^\dagger\mathbf{B}\mathbf{A}$, $\alpha_i$, are upper bounded $\alpha_i \leq \beta_i$. We therefore have
\begin{equation}
\label{eqn:detineq}
    \det \big( \mathbf{A}^\dagger\mathbf{B}\mathbf{A} \big) = \prod_{i = 1}^m \alpha_i \leq \prod_{i = 1}^m \beta_i.
\end{equation}
Setting the columns of $\mathbf{A}$ to be the $m$ principal eigenvectors of $\mathbf{B}$ achieves equality in (\ref{eqn:detineq}). This $\mathbf{A}$ is non-unique, since any $\mathbf{A}^\star = \mathbf{A}\mathbf{\Theta}$, where $\mathbf{\Theta} \in \mathbb{C}^{n \times n}$ is a unitary matrix, also achieves equality with the upper bound.


\subsection*{Appendix 2}
From standard information theory properties
\begin{equation}
\label{eqn:cond}
     \mathcal{I}(\mathbf{z}_1,\ldots,\mathbf{z}_L;\mathbf{x}) = \mathcal{I}(\mathbf{z}_{l}^{\mathsf{c}};\mathbf{x})
     + \mathcal{I}(\mathbf{z}_l;\mathbf{x}\vert \mathbf{z}_{l}^{\mathsf{c}}),
\end{equation}
Applying the matrix determinant lemma, $\det (\mathbf{A} + \mathbf{B}\mathbf{C}) = \det(\mathbf{I} + \mathbf{C}\mathbf{A}^{-1}\mathbf{B})\det(\mathbf{A}) $,
\begin{align}
\label{eqn:expansion}
\mathcal{I}(\mathbf{z}_1,\ldots,\mathbf{z}_L;\mathbf{x}) &= \log_2 \det \big( \mathbf{I}_K + \rho\sum_{i=1}^{l}\mathbf{H}_i^\dagger\mathbf{W}_i\mathbf{W}_i^\dagger\mathbf{H}_i\big) \nonumber \\ &= \log_2 \det \big( \mathbf{I}_{N} + \rho\mathbf{W}_l^\dagger\mathbf{H}_l\mathbf{A}_{l}\mathbf{H}_l^\dagger\mathbf{W}_l \big) + \log_2 \det \big(\mathbf{A}_{l}^{-1}\big)
\end{align}
where 
\begin{equation*}
    \mathbf{A}_{l} = \big(\mathbf{I}_K + \sum_{i \neq l}\mathbf{H}_i^\dagger\mathbf{W}_i\mathbf{W}_i^\dagger\mathbf{H}_i\big)^{-1}.
\end{equation*}
By inspection of (\ref{eqn:cond}) and (\ref{eqn:expansion})
\begin{equation*}
    \mathcal{I}(\mathbf{z}_l;\mathbf{x}\vert \mathbf{z}_{l}^{\mathsf{c}}) = \log_2 \det \big( \mathbf{I}_{N} + \rho\mathbf{W}_l^\dagger\mathbf{H}_l\mathbf{A}_{l}\mathbf{H}_l^\dagger\mathbf{W}_l \big).
\end{equation*}

\ifCLASSOPTIONcaptionsoff
  \newpage
\fi

\end{document}